\documentclass[aps,prb,preprint,superscriptaddress]{revtex4-1}

\usepackage{graphicx,amsmath,amsfonts,mathrsfs}

\begin{document}


\title{Magnetization induced local electric dipoles and multiferroic properties of Ba$_2$CoGe$_2$O$_7$}


\author{I. V. Solovyev}
\email{SOLOVYEV.Igor@nims.go.jp}
\affiliation{Computational Materials Science Unit,
National Institute for Materials Science,
1-1 Namiki, Tsukuba, Ibaraki 305-0044, Japan}
\affiliation{Department of Theoretical Physics and Applied Mathematics, Ural Federal University,
Mira str. 19, 620002 Ekaterinburg, Russia}


\date{\today}

\begin{abstract}
Ba$_2$CoGe$_2$O$_7$, crystallizing in the noncentrosymmetric but nonpolar $P\overline{4}2_1m$ structure,
belongs to a special class of multiferroic materials, whose properties are predetermined by the
availability of the
rotoinversion symmetry. Unlike inversion, the rotoinversion symmetry can be easily destroyed
by the magnetization. Moreover, due to specific structural pattern
in the $xy$ plane, in which the magnetic Co$^{2+}$ ions
are separated by the nonmagnetic GeO$_4$ tetraherda, the magnetic structure of Ba$_2$CoGe$_2$O$_7$
is relatively soft. Altogether, this leads to the rich variety of multiferroic properties of Ba$_2$CoGe$_2$O$_7$,
where the magnetic structure can be easily deformed by the magnetic field,
inducing the net electric polarization in the direction depending on the magnetic symmetry of the system,
which itself depends on the direction of the magnetic field.
In this paper, we show that all these properties can be successfully explained on the basis of
realistic low-energy model, derived from the first-principles electronic structure calculations
for the magnetically active Co $3d$ bands, and the Berry-phase theory of
electronic polarization. Particularly, we argue that the magnetization induced electric polarization
in Ba$_2$CoGe$_2$O$_7$
is essentially local and expressed via the expectation values
$\langle \hat{\boldsymbol{p}} \rangle = {\rm Tr} [ \hat{\boldsymbol{p}} \hat{\cal D} ]$
of some dipole matrices $\hat{\boldsymbol{p}}$, calculated in the Wannier basis of the model, and the
site-diagonal density matrices $\hat{\cal D}$ of the magnetic Co sites.
Thus, the basic aspects of the behavior of
Ba$_2$CoGe$_2$O$_7$ can be understood already in the atomic limit,
where both magnetic anisotropy and magnetoelectric coupling are
specified by $\hat{\cal D}$.
Then, the
macroscopic polarization
can be found as a superposition of $\langle \hat{\boldsymbol{p}} \rangle$ of the individual Co sites.
We discuss the behavior of interatomic magnetic interactions, main contributions to the
magnetocrystalline anisotropy and the spin canting in the $xy$ plane,
as well as the similarities and differences of the proposed picture from the
phenomenological model of spin-dependent $p$-$d$ hybridization.
\end{abstract}


\maketitle

\section{\label{sec:Intro} Introduction}

  Magnetically driven ferroelectricity is one of the major topics in the condensed matter physics today.
Even though the basic crystallographic symmetry of a magnetic material
may not allow for the existence of a
spontaneous electric polarization,
this symmetry can be further lowered by the magnetic order, which in some cases
gives rise to the ferroelectric (FE) activity.
After discovery of such effect in TbMnO$_3$,\cite{KimuraTbMnO3} where the inversion
symmetry is broken by some complex noncollinear magnetic order, there is a large number of experimental and
theoretical proposals pointing at the existence of a similar effect in other magnetic materials,
which are called multiferroics.\cite{MF_review}

\begin{figure}[h!]
\begin{center}
\includegraphics[width=10cm]{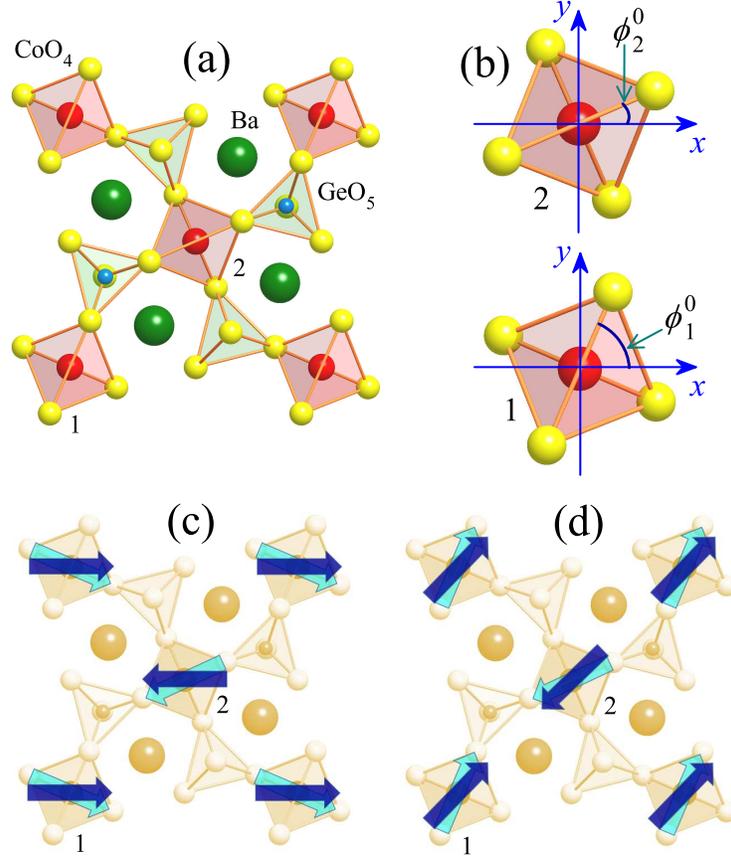}
\end{center}
\caption{\label{fig.structure}
(Color online)
Crystal and spin magnetic structure of Ba$_2$CoGe$_2$O$_7$: (a) General view on the crystal structure in the $xy$ plane.
The Co and O atoms are indicated by the medium red and yellow spheres, respectively, the Ba atoms are indicated by the
big green spheres, and the Ge atoms are indicated by the small blue spheres.
The MnO$_4$ and GeO$_4$ tetrahedra are colored red and green, respectively. (b)
Details of the rotations of MnO$_4$ tetrahedra
associated with two Mn sites
in the $xy$ plane. (c) and (d) Two possible spin magnetic structures, realized without external magnetic field.
The light (cyan) arrows indicate the directions of spins favored by the single-ion anisotropy, while the
dark (blue) arrows are the joint effect of the single-ion anisotropy and interatomic exchange interactions.
}
\end{figure}

  Ba$_2$CoGe$_2$O$_7$ and related compounds take a special place among multiferroic materials.
They crystallize in the noncentrosymmetric $P\overline{4}2_1m$ structure,\cite{Hutanu11} which nonetheless
does not permit the FE effect because of the fourfold rotoinversion symmetry.
Nevertheless, unlike inversion,
the rotoinversion symmetry can be relatively easy destroyed by the magnetic order, by canting
the magnetic moments out of the rotoinversion axis. Thus, one unique aspect of Ba$_2$CoGe$_2$O$_7$ is that
the ferroelectricity in this compound
can be induced relatively simple C-type antiferromagnetic (AFM) order.
Furthermore, the magnetic structure of Ba$_2$CoGe$_2$O$_7$
is relatively soft: It consists of the CoO$_4$ tetrahedra, which are interconnected with each other via
nonmagnetic GeO$_4$ tetrahedra (see Fig.~\ref{fig.structure}a).
Since two magnetic Co sites in this structure are separated by the long Co-O-Ge-O-Co paths, the exchange interactions
between them are expected to be small and the magnetic structure can be easily deformed by the external
magnetic field. The possibilities of easy manipulation by the magnetic structure and
switching the FE properties
have attracted a great deal of attention to Ba$_2$CoGe$_2$O$_7$, and
today it was demonstrated in many details how the electric polarization in Ba$_2$CoGe$_2$O$_7$
can be tuned by the magnetic field (Refs.~\onlinecite{Yi,Murakawa})
as well as the uniaxial stress (Ref.~\onlinecite{Nakajima}).

  The experimental behavior of the electric polarization in Ba$_2$CoGe$_2$O$_7$
is frequently interpreted basing
on the model model of spin-dependent $p$-$d$ hybridization:
if $\boldsymbol{e}$ is the direction of the spin magnetic moment at the Co site,
located in the origin, and $\boldsymbol{\upsilon}$ is the position of a ligand oxygen atom,
relative to this origin, the bond Co-O contributes to the local electric dipole
moment as
$\boldsymbol{p} \propto (\boldsymbol{e} \cdot \boldsymbol{\upsilon})^2 \boldsymbol{\upsilon}$.\cite{Murakawa,Arima}
This simple phenomenological expression
is able to capture the symmetry properties of the electric polarization and, thus,
explain the behavior of this polarization on a phenomenological level.
It is therefore not surprising that a qualitatively similar behavior of the electric polarization
was obtained in the first-principles electronic structure calculations,\cite{Yamauchi}
which generally support the model of spin-dependent $p$-$d$ hybridization.

  On the other hand, the correct quantum-mechanical definition of the electric polarization is solids
should be based on the Berry-phase theory.\cite{FE_theory} This theory provides not only
an efficient computational framework, which is used today in most of the first-principles electronic
structure calculations, but also appears to be a good starting point for the construction of
realistic microscopic models,
explaining the origin and
basic aspects of the behavior of electric polarization in various types of compounds.\cite{Barone,DEPol,Terakura}
In this work we will
pursue this strategy for the analysis of
electric polarization in Ba$_2$CoGe$_2$O$_7$.
First, we will show that, in the atomic (Wannier) basis, the electric polarization
consists of two parts: the local one, which is expressed via the expectation value
$\langle \hat{\boldsymbol{p}} \rangle = {\rm Tr} [ \hat{\boldsymbol{p}} \hat{\cal D} ]$
of some local dipole matrices $\hat{\boldsymbol{p}}$ and
site-diagonal density matrices $\hat{\cal D}$, and the anomalous one, which is solely related
to the phases of the coefficients of the expansion of the wavefunctions in the basis of
atomic Wannier orbitals.
Similar classification holds for the orbital magnetization.\cite{Nikolaev}
Then, if the crystal structure possesses the inversion symmetry,
the local term becomes inactive and the electric polarization, induced by the magnetic inversion
symmetry breaking, is anomalous in origin.
Such situation occurs, for instance, in multiferroic manganites with the orthorhombic structure.\cite{DEPol,orth_MF}
In Ba$_2$CoGe$_2$O$_7$, however, the situation is exactly the opposite: to a good approximation,
the electric polarization
is expressed as a sum of local electric dipole moments
of individual Co sites, while the anomalous contribution is negligibly small.
This explains many experimental details of the behavior of electric polarization in Ba$_2$CoGe$_2$O$_7$
as well as
basic difference of this compound from other multiferroic materials.

  The rest of the paper is organized as follows. The method is described in Sec.~\ref{sec:Method}.
We use the same strategy as in our previous
works, devoted to the analysis of multiferroic properties of various transition-metal oxides
on the basis of effective realistic low-energy models, derived from the first-principles calculations.\cite{DEPol,orth_MF}
Therefore, in Sec.~\ref{sec:LEmodel} we will remind the main details of the construction of such
effective low-energy model. In Sec.~\ref{sec:EPol}, we will analyze the main contributions
to the electric polarization in the case of basis, when the Bloch wavefunction
is expanded over some atomic Wannier orbitals, centered at
the magnetic sites.
Since we do not consider the magnetostriction effects, which can low the original $P\overline{4}2_1m$
symmetry of the crystal structure, the considered electric polarization is in fact the electronic one.\cite{FE_theory}
In Sec.~\ref{sec:Results} we will present results of our calculations for Ba$_2$CoGe$_2$O$_7$,
starting from the atomic limit and, then, consecutively considering the effect of interatomic exchange interactions
and the external magnetic field.
Finally, in Sec.~\ref{sec:Summary}, we will summarize our work. In the Appendix we will estimate
the ferromagnetic (FM) contribution to the interatomic exchange interaction cause by the magnetic polarization
of the oxygen $2p$ band and discuss how such effect can be evaluated starting from the low-energy
electron model for the magnetic Co $3d$ bands.

\section{\label{sec:Method} Method}

\subsection{\label{sec:LEmodel} Effective low-energy model}

  In this section, we briefly remind the reader
the main ideas behind the construction of the effective low-energy electron model.
The details can be found in the review article (Ref.~\onlinecite{review2008}).

\begin{figure}[h!]
\begin{center}
\includegraphics[width=10cm]{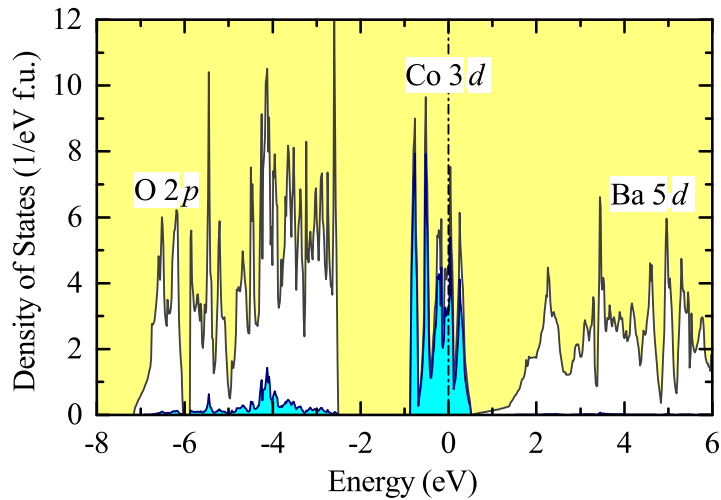}
\end{center}
\caption{\label{fig.DOS}
(Color online) Total and partial densities of states of Ba$_2$CoGe$_2$O$_7$ in the local density approximation.
The shaded light (blue) area shows the contribution of the Co $3d$ states.
The positions of the main bands are indicated by symbols. The Fermi level is at zero energy (shown by dot-dashed line).
}
\end{figure}

  The model Hamiltonian,
\begin{equation}
\hat{\cal{H}}  =  \sum_{ij} \sum_{\alpha \beta}
t_{ij}^{\alpha \beta}\hat{c}^\dagger_{i\alpha}
\hat{c}^{\phantom{\dagger}}_{j\beta} +
  \frac{1}{2}
\sum_{i}  \sum_{\alpha \beta \gamma \delta} U^i_{\alpha \beta
\gamma \delta} \hat{c}^\dagger_{i\alpha} \hat{c}^\dagger_{i\gamma}
\hat{c}^{\phantom{\dagger}}_{i\beta}
\hat{c}^{\phantom{\dagger}}_{i\delta},
\label{eqn.ManyBodyH}
\end{equation}
is formulated
in the basis of Wannier orbitals $\{ \phi_{i \alpha} \}$,
which are constructed for the magnetically active Co $3d$ bands near the Fermi level,
starting from the band structure in the local-density approximation (LDA)
without spin-orbit (SO) coupling (Fig.~\ref{fig.DOS}).
Here, each Greek symbol ($\alpha$, $\beta$, $\gamma$, or $\delta$) stands for the combination of
spin ($\sigma_{\alpha}$, $\sigma_{\beta}$, $\sigma_{\gamma}$, or $\sigma_{\delta}$) and orbital ($a$, $b$, $c$, or $d$) indices,
for which we adapt the following order: $xy$, $yz$, $3z^2$$-$$r^2$, $zx$, $x^2$$-$$y^2$.
Each lattice point $i$ ($j$) is specified by the position
$\boldsymbol{\tau}$ ($\boldsymbol{\tau}'$)
of the atomic site in the primitive cell and the lattice translation ${\bf R}$.
Hence, the basis orbitals
$\phi_{i \alpha}({\bf r}) \equiv \phi_{\tau \alpha} ({\bf r}$$-$${\bf R}$$-$$\boldsymbol{\tau})$
are centered in
the lattice point $({\bf R}$$+$$\boldsymbol{\tau})$
and labeled by the indices $\tau$ and $\alpha$.
Moreover, they satisfies the orthonormality condition:
\begin{equation}
\langle\phi_{\tau'\alpha'}(\textbf{r}-\textbf{R}'-\boldsymbol{\tau}')|
\phi_{\tau \alpha}(\textbf{r}-\textbf{R}-\boldsymbol{\tau})\rangle=
\delta_{\textbf{R}'\textbf{R}} \delta_{\tau' \tau} \delta_{\alpha'\alpha}.
\label{eqn:orthogon}
\end{equation}

  In our case, the Wannier function were calculated using the
projector-operator method (Refs.~\onlinecite{review2008,WannierRevModPhys})
and orthonormal linear muffin-tin orbitals (LMTO's, Ref.~\onlinecite{LMTO}) as the trial wave functions.
Typically such procedure allows us to generate
well localized Wannier functions, that is guaranteed by the good localization of LMTO's themselves.
Then, the one-electron part of the model (\ref{eqn.ManyBodyH})
is identified with the matrix elements of LDA Hamiltonian (${\cal H}_{\rm LDA}$) in the Wannier basis:
$t^{\alpha \beta}_{\boldsymbol{\tau}, \boldsymbol{\tau}'+{\bf R}} =
\langle \phi_{\tau \alpha} ({\bf r}$$-$$\boldsymbol{\tau})| {\cal H}_{\rm LDA} |
\phi_{\tau' \beta} ({\bf r}$$-$${\bf R}$$-$$\boldsymbol{\tau}') \rangle$.
Since the Wannier basis is complete in the low-energy part of the spectrum, the construction is exact in the sense that
the band structure, obtained from $t^{\alpha \beta}_{\boldsymbol{\tau}, \boldsymbol{\tau}'+\bf R}$,
exactly coincides with the one of LDA.

  All Wannier basis and one-electron parameters $t^{\alpha \beta}_{\boldsymbol{\tau}, \boldsymbol{\tau}'+{\bf R}}$ were
first computed without the SO interaction.
Then, the SO interaction at each atomic site was included in the ``second-variation step'', in the
basis of the nonrelativistic Wannier functions:
$\langle \phi_{i \alpha} ({\bf r}) | \Delta {\cal H}_{\rm SO} | \phi_{i \beta} ({\bf r}) \rangle$,
as explained in Ref.~\onlinecite{review2008}.

  Matrix elements of screened Coulomb interactions can be also computed in the Wannier basis as
$$
U_{\alpha \beta \gamma \delta}^i = \int d {\bf r} \int d {\bf r}'
\phi_{i \alpha}^* ({\bf r})
\phi_{i \beta} ({\bf r}) v_{\rm scr}({\bf r},{\bf r}')
\phi_{i \gamma}^* ({\bf r}') \phi_{i \delta} ({\bf r}'),
$$
where
$v_{\rm scr}({\bf r},{\bf r}')$ is obtained using the constrained RPA technique.\cite{Ferdi04}
Then, $v_{\rm scr}({\bf r},{\bf r}')$ does not depend on spin variables and
$U_{\alpha \beta \gamma \delta}^i = U_{abcd}^i \,
\delta_{\sigma_\alpha \sigma_\beta} \delta_{\sigma_\gamma \sigma_\delta}$.
Since RPA is very time consuming technique, we employ additional simplifications, which were discussed
in Ref.~\onlinecite{review2008}. Namely, first we evaluate the screened Coulomb and exchange interactions between
atomic Co $3d$ orbitals, using fast and more suitable for these purposes constrained LDA technique. After that,
we consider additional channel of screening caused by the $3d \rightarrow 3d$ transitions
in the polarization function of constrained RPA and projecting this
function onto atomic $3d$ orbitals. The so obtained parameters of screened Coulomb interactions
are well consistent with results of full-scale constrained RPA calculations
without additional simplifications.

  All calculations have been performed for the room temperature $P\overline{4}2_1m$ structure
reported in Ref.~\onlinecite{Hutanu11}. The corresponding parameters of the low-energy
model are summarized in supplemental materials.\cite{SM}

  After the construction, the model is solved in the unrestricted Hartree-Fock (HF) approximation,
which is well justified for the considered case where the degeneracy of the ground state is lifted
by the crystal distortion.\cite{review2008}

\subsection{\label{sec:EPol} Electronic polarization}

  The electronic polarization can be
computed in the reciprocal space,
using the formula of King-Smith and Vanderbilt:\cite{FE_theory}
\begin{equation}
\textbf{P} = - \frac{ie}{(2 \pi)^3} \sum_{n}
\int_{BZ} d \textbf{k} \,  \langle u_{n \textbf{k}} | \partial_{\textbf{k}} u_{n \textbf{k}} \rangle ,
\label{eqn:PKSV}
\end{equation}
where $u_{n \textbf{k}}(\textbf{r}) = e^{-i \textbf{kr}} \psi_{n \textbf{k}}(\textbf{r})$ is the cell-periodic eigenstate
of the Hamiltonian $H_{\textbf{k}} = e^{-i \textbf{kr}} H e^{i \textbf{kr}}$, the summation runs over the
occupied bands ($n$), the $\textbf{k}$-space integration goes over the first
Brillouin zone (BZ), and $-$$e$ ($e > 0$) is the electron charge. In our case, each
$\psi_{n \textbf{k}} (\textbf{r})$ is expanded in the basis of Wannier orbitals
$\phi_{\tau \alpha} (\textbf{r}-\textbf{R}-\boldsymbol{\tau})$, used for the construction of the low-energy model:
\begin{equation}
\psi_{n\textbf{k}} (\textbf{r}) = \frac{1}{\sqrt{N}}
\sum_{\textbf{R} \tau \alpha} c^{\tau \alpha}_{n\textbf{k}}
e^{i\textbf{k}(\textbf{R}+\boldsymbol{\tau})} \phi_{\tau \alpha}(\textbf{r}-\textbf{R}-\boldsymbol{\tau}),
\label{eqn:lmtofunc}
\end{equation}
where $N$ is the number of primitive cells.
Then, the
\textbf{k}-space gradient of $u_{n\textbf{k}}$ will have two contributions:
\begin{equation}
\begin{aligned}
\partial_{\textbf{k}}u_{n\textbf{k}} = & -\frac{i}{\sqrt{N}}
\sum_{\textbf{R} \tau \alpha}(\textbf{r}-\textbf{R}-\boldsymbol{\tau})e^{-i\textbf{k}(\textbf{r}-\textbf{R}-\boldsymbol{\tau})}
c^{\tau \alpha}_{n\textbf{k}}
\phi_{\tau \alpha}(\textbf{r}-\textbf{R}-\boldsymbol{\tau})
\\
&+\frac{1}{\sqrt{N}}
\sum_{\textbf{R} \tau \alpha}
e^{-i\textbf{k}(\textbf{r}-\textbf{R}-\boldsymbol{\tau})}
\partial_{\textbf{k}}c^{\tau \alpha}_{n\textbf{k}} \phi_{\tau \alpha}(\textbf{r}-\textbf{R}-\boldsymbol{\tau}),
\label{eqn:uder}
\end{aligned}
\end{equation}
and the electronic polarization \textbf{P} will also include two terms:
\begin{equation}
\textbf{P} = \sum_{n}
\int_{BZ} \frac{d\textbf{k}}{\Omega} \,
\langle c_{n \textbf{k}} | \hat{\boldsymbol{p}}_{\textbf{k}} | c_{n \textbf{k}} \rangle
- \frac{ie}{(2 \pi)^3} \sum_{n}
\int_{BZ} d\textbf{k} \,
\langle c_{n \textbf{k}} | \partial_{\textbf{k}} c_{n \textbf{k}} \rangle.
\label{eqn:Pbasis}
\end{equation}
Here, $| c_{n \textbf{k}} \rangle$ denotes the column vector
$| c_{n \textbf{k}} \rangle \equiv [c^{\tau \alpha}_{n \textbf{k}}]$,
$\hat{\boldsymbol{p}}_{\textbf{k}}$ is the matrix
$\hat{\boldsymbol{p}}_{\textbf{k}} \equiv [\boldsymbol{p}^{\tau' a',\tau a}_{\textbf{k}}
\delta_{\sigma_{\alpha'}\sigma_{\alpha\phantom{'}}}]$, where
\begin{equation}
\boldsymbol{p}^{\tau' a',\tau a}_{\textbf{k}} =
- \frac{e}{V} \sum_{\textbf{R}} \langle \phi_{\tau' \alpha'}(\textbf{r}-\boldsymbol{\tau}') | \textbf{r} |
\phi_{\tau \alpha}(\textbf{r}-\textbf{R}-\boldsymbol{\tau})
\rangle e^{i \textbf{k}(\textbf{R}+\boldsymbol{\tau}-\boldsymbol{\tau}')},
\label{eqn:pmatrix}
\end{equation}
$V$ is the unit cell volume, and $\Omega = (2 \pi)^3/V$ is the first BZ volume.
Moreover, since $\textbf{P}$ is understood as the \textit{change} of the polarization in the process of adiabatic
symmetry lowering (Ref.~\onlinecite{FE_theory}),
which in our case is driven by the magnetic degrees of freedom,
here and below we drop all
nonmagnetic contributions to $\boldsymbol{p}^{\tau' a',\tau a}_{\textbf{k}}$,
which are irrelevant to the magnetic symmetry lowering.

  If the transition-metal site is located in the inversion center, the first term of Eq.~(\ref{eqn:Pbasis})
identically vanishes.
Such situation is realized, for instance, in multiferroic manganites, crystallizing in the orthorhombic $Pbnm$ structure,
where the FE activity is entirely ``anomalous'' and associated with the second term of Eq.~(\ref{eqn:Pbasis}).\cite{orth_MF}
However, in Ba$_2$CoGe$_2$O$_7$ the situation appears to be exactly the opposite:
the first term dominates, while the second contribution is negligibly small
(about $0.1$ $\mu$C/m$^2$ in the magnetic ground state).
Furthermore, due to the orthogonality condition (\ref{eqn:orthogon}),
the leading contribution to
$\boldsymbol{p}^{\tau' a',\tau a}_{\textbf{k}}$ comes from the
Wannier functions centered at the same atomic site. Then, one can impose in Eq.~(\ref{eqn:pmatrix}) the additional
condition $\boldsymbol{\tau}' = \textbf{R}$$+$$\boldsymbol{\tau}$, which yields
$\boldsymbol{p}^{\tau' a',\tau a}_{\textbf{k}} \approx
\boldsymbol{p}^{a' a}_{\tau} \delta_{\tau' \tau}$,
where $\boldsymbol{p}^{a' a}_{\boldsymbol{\tau}}$ does not depend on \textbf{k}.
Thus, the electronic polarization in Ba$_2$CoGe$_2$O$_7$ will be given by the sum
\begin{equation}
\textbf{P} \approx \sum_{\tau} \textbf{P}_{\tau}
\label{eqn:papprox}
\end{equation}
of the local electric dipoles:
\begin{equation}
\textbf{P}_{\tau} \equiv \langle \hat{\boldsymbol{p}}_{\tau} \rangle
= {\rm Tr}_a \{ \hat{\cal D}_{\tau} \hat{\boldsymbol{p}}_{\tau} \},
\label{eqn:Ldipole}
\end{equation}
where ${\rm Tr}_a$ is the trace over $a$.
Each such dipole is given by the expectation value $\langle \hat{\boldsymbol{p}}_{\tau} \rangle$ of the dipole matrix
$\hat{\boldsymbol{p}}_{\tau} = [\boldsymbol{p}^{a' a}_{\tau}]$
and the spin-independent part of the density matrix $\hat{\cal D}_{\tau} = [{\cal D}^{\alpha \alpha'}_{\tau}]$
at the site $\tau$:
$$
{\cal D}^{a a'}_{\tau} = \sum_{\sigma_{\alpha}} \delta_{\sigma_{\alpha\phantom{'}} \sigma_{\alpha'}}
\sum_{n} \int_{BZ}
\, \frac{d\textbf{k}}{\Omega} \,c^{\tau \alpha}_{n\textbf{k}} c^{\tau \alpha'*}_{n\textbf{k}}.
$$

  In order to evaluate $\hat{\boldsymbol{p}}_{\tau}$
we use the LMTO method and expand
each $\phi_{\tau \alpha}(\textbf{r})$ is the basis $\{ \chi_{\upsilon \beta} \}$
of linear muffin-tin orbitals,
which can be viewed as orthonormalized atomic-like Wannier orbitals,
constructed in the whole region of valence states.\cite{LMTO}
Moreover, we shift each site $\tau$ to the origin, that does not affect the polarization change
caused by the magnetic symmetry lowering.
Then, without SO interaction, we have
\begin{equation}
\phi_{\tau a}(\textbf{r}) = \sum_{\upsilon b}
q_{\tau a}^{\upsilon b} \chi_{\upsilon b} (\textbf{r} - \boldsymbol{\upsilon} + \boldsymbol{\tau}).
\label{eqn:overLMTO}
\end{equation}
where,
$\upsilon = \tau$ corresponds to the ``head'' of the
Wannier function, centered at the Co-site $\tau$, while all other contributions describe the ``tails''
of the Wannier functions, spreading to the
neighboring sites $\upsilon$. Then, around each such site we identically present the position operator as
$\textbf{r} = (\boldsymbol{\upsilon}$$-$$\boldsymbol{\tau}) + (\textbf{r}$$-$$\boldsymbol{\upsilon}$$+$$\boldsymbol{\tau})$
and assume that the leading contribution to $\hat{\boldsymbol{p}}_{\tau}$
comes from the first term.
This is reasonable because, at each site, $(\textbf{r} - \boldsymbol{\upsilon} + \boldsymbol{\tau})$ couples the
atomic states with different parity, which are typically well separated in energy.
Then, the matrix elements $\boldsymbol{p}^{\alpha \alpha'}_{\tau}$ can be easily evaluated as
$$
\boldsymbol{p}^{a'a}_{\tau} = -\frac{e}{V} \sum_{\upsilon b}
(\boldsymbol{\upsilon} - \boldsymbol{\tau}) q_{\tau a'}^{\upsilon b *}
q_{\tau a}^{\upsilon b}.
$$

  Then, by considering the leading contributions of four oxygen sites surrounding each Co atom,
one obtains the following matrices
(in $\mu$C/m$^2$):
\begin{equation}
\hat{p}_{1,2x} = \left(
\begin{array}{ccccc}
              0 & \mp 3509 &       0 & \phantom{-}6357 &        0 \\
       \mp 3509 &        0 &    -801 &               0 &    -3553 \\
              0 &     -801 &       0 &         \pm 303 &        0 \\
\phantom{-}6357 &        0 & \pm 303 &               0 & \mp 6142 \\
              0 &    -3553 &       0 &        \mp 6142 &        0 \\
\end{array}
\right),
\label{eqn:p12x}
\end{equation}
\begin{equation}
\hat{p}_{1,2y} = \left(
\begin{array}{ccccc}
              0 & \phantom{-}6357 &       0 &        \pm 3509 &               0 \\
\phantom{-}6357 &               0 & \mp 303 &               0 &        \mp 6142 \\
              0 &         \mp 303 &       0 &            -801 &               0 \\
       \pm 3509 &               0 &    -801 &               0 & \phantom{-}3553 \\
              0 &        \mp 6142 &       0 & \phantom{-}3553 &               0 \\
\end{array}
\right),
\label{eqn:p12y}
\end{equation}
and
\begin{equation}
\hat{p}_{1,2z} = \left(
\begin{array}{ccccc}
    0 &               0 &           -3682 &               0 &        0 \\
    0 &        \pm 3394 &               0 & \phantom{-}3524 &        0 \\
-3682 &               0 &               0 &               0 & \pm 2870 \\
    0 & \phantom{-}3524 &               0 &        \mp 3394 &        0 \\
    0 &               0 &        \pm 2870 &               0 &        0 \\
\end{array}
\right),
\label{eqn:p12z}
\end{equation}
where the upper (lower) signs correspond to the Co-sites 1 (2) in Fig.~\ref{fig.structure}. As expected, the matrices
$\boldsymbol{p}_{1,2} \equiv (\hat{p}_{1,2x},\hat{p}_{1,2y},\hat{p}_{1,2z})$ obey the $P\overline{4}2_1m$ symmetry.
If the density matrix $\hat{\cal D}_{1,2}$ obeys the same symmetry, all local electric dipoles will vanish and
there will be no net polarization in the ground state.
However, if the symmetry of $\hat{\cal D}_{1,2}$ is lowered by magnetism, one can expect the
appearance of local electric dipoles with some order between sites 1 and 2. If this order is antiferroelectric,
there will be no net polarization. Nevertheless, if this order permits a FE component,
the system will exhibit a finite net polarization. The details of such symmetry lowering will depend
on directions of magnetic moments at the sites 1 and 2. For instance, if the spins are parallel to the
$z$ axis, the fourfold rotoinversion around $z$, $\hat{S}_4^z$, will remain among the symmetry
operations of the magnetic space group,
and there will be no local electric dipoles. However, if the moments
lie in the $xy$ plane, the rotoinversion $\hat{S}_4^z$ is replaced by the regular twofold rotation, $\hat{C}_2^z$,
which allows for the existence of local electric dipoles parallel to $z$. For an arbitrary direction of spin,
the symmetry will be further lowered, and the local electric dipoles may have all three components. However,
the relative directions of dipoles at the sites 1 and 2 will depend on other symmetry operations.
Below, we will investigate this magnetic symmetry lowering more in details.

  To conclude this section, we would like to emphasize that Eqs.~(\ref{eqn:papprox}) and (\ref{eqn:Ldipole}) are the
correct quantum-mechanical expressions for the electric polarization in Ba$_2$CoGe$_2$O$_7$, which are
based on the very general Berry-phase theory.\cite{FE_theory} These expressions allows us clarify the
microscopic origin of the polarization, which is frequently ascribed to the
spin-dependent $p$-$d$ hybridization.\cite{Yamauchi}
The $p$-$d$ hybridization does play a very important role in this material as it defines
the matrix elements of $\hat{\boldsymbol{p}}_{\tau}$ in Eq.~(\ref{eqn:Ldipole}).
Moreover, the contribution of each Co-O bond
to the local electric dipole moment is indeed proportional to
the vector $(\boldsymbol{\upsilon}$$-$$\boldsymbol{\tau})$, connecting the Co and O sites.
However,
the matrix elements ${\boldsymbol{p}}_{\tau}^{a'a}$
themselves do not depend on the spin state. In this sense, the spin dependence
of the hybridization is not the most important aspect.
The spin dependence of the electric polarization in our picture
comes from the matrix elements of the density matrix, which describes the effect of the SO coupling
at the transition-metal sites. The local magnetic moment deforms the electron density around the Co sites.
This deformation, which itself depends on the direction of local magnetic moment,
spreads to the neighboring oxygen sites via the tails of the Wannier functions
and
produces finite
electric moment. This is the basic microscopic picture underlying the formation of the
local electric moments in Ba$_2$CoGe$_2$O$_7$.

\section{\label{sec:Results} Results and Discussions}

  The first important question we need to address is the local properties developed at each of the Co sites.
For these purposes we set all transfer integrals in Eq.~(\ref{eqn.ManyBodyH}) equal to zero and solve the model in the
mean-field approximation separately for each Co site. The SO interaction in these calculations is treated
in the frameworks of the self-consistent linear response (SCLR) theory.\cite{SCLR} For each direction of spin,
it gives us the self-consistent HF potential in the first order of the SO coupling, which can be used again
as the input of HF equations in order to obtain at the output the change of
the density matrix and the total energy beyond the first order. We would like to emphasize that
the considered below effects are beyond the first order of the SO coupling.

\begin{figure}[h!]
\begin{center}
\includegraphics[width=10cm]{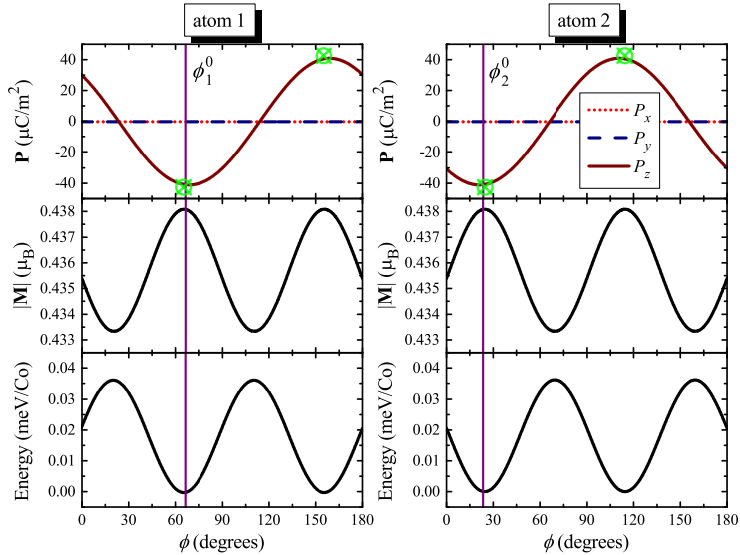}
\end{center}
\caption{\label{fig.atomicP}
(Color online)
Electronic polarization, absolute value of orbital magnetization, and total energy
depending on the direction
$\boldsymbol{e} = (\cos \phi , \sin \phi , 0)$ of spin magnetization in the $xy$ plane.
The lines show results of self-consistent linear response theory for the
spin-orbit coupling.\cite{SCLR} The symbols show results of unrestricted Hartree-Fock calculations
for the high-symmetry points.
}
\end{figure}

  The obtained polarization is in the good agreement with the regular
self-consistent HF calculations, which can be performed for the high-symmetry points.
For the in-plane rotations of the spin magnetization, the results are summarized in Fig.~(\ref{fig.atomicP}).
In this case, $P_{x}$ and $P_{y}$
identically vanish, while $P_{z}(\phi)$ exhibits the characteristic cosine-like behavior.\cite{Murakawa}
As expected, each $P_{z}(\phi)$ takes its minimum value at the angle $\phi^0_{\tau}$, which specifies the
direction of the upper O-O bond of the CoO$_4$ tetrahedron (see Fig.~\ref{fig.structure}).
Note also the minus sign in Eq.~(\ref{eqn:pmatrix}), which means that the considered polarization
is purely electronic.
Since two Co sites
in Ba$_2$CoGe$_2$O$_7$ are connected by the symmetry operation $\{\hat{C}_2^x | \boldsymbol{a}_1/2$$+$$\boldsymbol{a}_2/2 \}$
(the twofold rotation around $x$, $\hat{C}_2^x$, associated with the half of the primitive translations in the $xy$ plane),
the angles $\phi^0_2$ and $\phi^0_1$ satisfy the property: $\phi^0_2 = \pi/2 - \phi^0_1$, where the experimental value of
$\phi^0_1$ is $67^\circ$.\cite{Hutanu11} The maximums of $P_{z}(\phi)$ are at $\phi = \phi^0_{\tau} + \pi/2$,
which specify the directions of the lower O-O bonds of the CoO$_4$ tetrahedra.
Correspondingly, $P_{z}(\phi) = 0$ when the spin moment is aligned in between the upper and lower O-O bonds
($\phi = \phi^0_{\tau} + \pi/4$), and the negative and positive contributions to the polarization
cancel each other.

  Very generally, the single-ion (SI) anisotropy energy in
tetragonal systems has the following form:\cite{Skomski}
$$
E_{\rm SI} = K_1 \sin^2 \theta + K_2 \sin^4 \theta + K_2' \sin^4 \theta \cos 4 (\phi - \phi^0_{\tau}).
$$
By fitting the total energies, obtained in the SCLR calculations, one can find that
$K_1 = -2.307$ meV, $K_2 = 0.072$ meV, and $K_2' = -$$0.018$ meV.
The obtained value of $K_1$ is well consistent with the experimental estimate $K_1 \approx -2.327$ meV,
reported in Refs.~\onlinecite{Soda,note2}. The parameter $K_2'$, controlling the in-plane anisotropy, is small but finite.
Anyway, this behavior is
different from the conventional $S=3/2$ spin-only model, where \textit{no} in-plane anisotropy
is expected.\cite{Soda} The difference is caused by an appreciable orbital magnetization
($\sim 0.434$ $\mu_{\rm B}$ per Co site), which is unquenched in the $xy$ plane. Moreover, this orbital
magnetization exhibits a sizable anisotropy ($\sim$$0.005$ $\mu_{\rm B}$ per Co site -- see Fig.~\ref{fig.atomicP}),
depending on the direction in the $xy$ plane. Then, it is reasonable to expect that the
magnetocrystalline anisotropy energy should be related to the behavior
of the orbital magnetization.\cite{Bruno} Indeed, we observe a close correlation between behavior
of the SI anisotropy energy and the anisotropy of the orbital magnetization
in the $xy$ plane (see Fig.~\ref{fig.atomicP}). Thus, similar to the polarization,
the origin of the in-plane SI anisotropy is related to the orbital degrees of freedom, which
become active due to the magnetic symmetry lowering. The SI anisotropy can be further controlled by
applying the uniaxial stress.\cite{Nakajima}

  By summarizing the behavior of the SI anisotropy energy, the spin moments are expected
to lie in the $xy$ plane and, at each Co site, be parallel to either upper or lower O-O bond.
Therefore, as far as the SI anisotropy energy is concerned, one expects two magnetic configurations,
which are depicted in Figs.~\ref{fig.structure}c and \ref{fig.structure}d. In the first case (Fig.~\ref{fig.structure}c),
the spins at the site 1 and 2 are parallel to, respectively,
lower and upper O-O bonds and hence this phase if antiferroelectric.
In the second case (Fig.~\ref{fig.structure}d), all spins are parallel to the upper bonds,
giving rise to the FE order. In the atomic limit,
these two configurations are degenerate.

\begin{figure}[h!]
\begin{center}
\includegraphics[width=6cm]{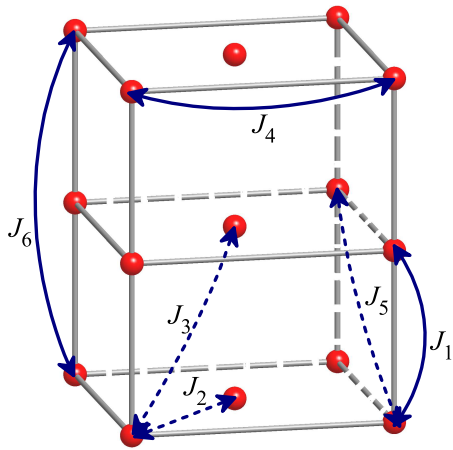}
\end{center}
\caption{\label{fig.J}
(Color online)
Main parameters of isotropic exchange interactions.
}
\end{figure}

  Another factor, controlling the relative direction of spins, is the interatomic exchange interactions.
The isotropic part of these interactions can be computed by considering the infinitesimal rotations of spins,
which provides the local mapping of the total energies onto the spin (Heisenberg) model
$E_{\rm H} = -\sum_{i>j} J_{ij} \boldsymbol{e}_i \cdot \boldsymbol{e}_j$, with $\boldsymbol{e}_i$
denoting the \textit{direction of spin} at the site $i$.\cite{JHeisenberg} There are six types of
nonvanishing exchange interactions, which are explained in Fig.~\ref{fig.J}.
The values of these interactions, obtained in the C-type AFM state, are
$J_1 = 0.012 $ meV, $J_2 = -$$2.730 $ meV, $J_3 = -$$0.046$ meV, $J_4 = -$$0.071$ meV, $J_5 = -$$0.007$ meV, and $J_6 = -$$0.012$ meV.
Thus, the exchange interaction $J_2 \equiv J$ stabilizes the AFM coupling in the $xy$ plane,
while the FM
coupling between the planes is stabilized by the combination of $J_1$, $J_3$, and $J_5$.
The leading interaction $J$ is relatively weak (compared to other transition-metal oxides),
that is directly related with
the crystal structure of Ba$_2$CoGe$_2$O$_7$, where neighboring Co sites in the $xy$ plane are separated by relatively long
Co-O-Ge-O-Co paths. Nevertheless, the value of $J$ seems to be overestimated in comparison with the experimental data.
For example,
the N\'eel temperature $T_{\rm N} \approx 34$ K,
evaluated using Tyablikov's random-phase approximation,\cite{spinRPA}
is about five time larger than the experimental $T_{\rm N} = 6.7$ K.\cite{Hutanu11}
Similar disagreement is found for
the exchange coupling itself: our calculations overestimate
the experimental $J$, reported in
Refs.~\onlinecite{Hutanu14,Soda,note2}, by the same order of magnitude.
This seems to be a negative aspect of the low-energy model (\ref{eqn.ManyBodyH}),
which was constructed only for the Co $3d$ bands and neglects several important contributions
to the magnetic properties of Ba$_2$CoGe$_2$O$_7$, related
to the magnetic polarization of the O $2p$ band. This is also the main reason why
we are not able to obtain a good quantitative agreement with the experimental data
for the behavior of electric polarization in the magnetic field: although our low-energy
model correctly reproduces the main tendencies, the magnetic field should by additionally
scaled (by the same factor as the exchange coupling $J$) to be compared with the
experimental data.
In the Appendix, we will evaluate the change of the magnetic energy, caused by the
polarization of the O $2p$ band, and show that this mechanism, which favors the FM alignment,
can indeed reduce the effective AFM coupling $J$. Nevertheless, this mechanism overestimates
the FM contribution to $J$ and alone does not resolve the quantitative disagreement
with the experimental data, which should involve additional factors, such as the
exchange striction below $T_{\rm N}$.

  The Dzyaloshinskii-Moriya (DM) interactions can be calculated by applying SCLR theory
for the SO coupling and considering a mixed perturbation, where the SO interaction is
combined with rotations of the spin magnetization.\cite{SCLR} The DM interactions between nearest neighbors along $z$
are forbidden by the symmetry. Therefore, the strongest interaction is again $\boldsymbol{d}_2 \equiv (d_x,d_y,d_z)$ , which takes place between
nearest neighbors in the $xy$ plane. For the bond, connecting the sites 1 and 2 in Fig.~\ref{fig.structure}a,
this vector is given by $\boldsymbol{d}_2 = (-$$5, 5, -$$6)$ $\mu$eV.
Similar parameters for other bonds can be obtained by applying
the symmetry operations of the space group $P\overline{4}2_1m$, which will change the signs of $d_x$ and $d_y$.
Thus, $\boldsymbol{d}_2$ is comparable with $K_2'$ and can also contribute to the canting of spins.

  The AFM interaction $J$ enforces the collinear alignment between spins in the $xy$ plane (see Fig.~\ref{fig.structure}),
while the SI anisotropy and DM interactions result in a small canting of spins, which is given by:\cite{SM}
\begin{equation}
\delta \phi \approx \frac{K_2' \cos 4 \Phi \sin 4\phi^0_1 + d_z}{2J},
\label{eqn:canting}
\end{equation}
where $\Phi = \frac{1}{2}(\phi_1 + \phi_2)$ is the average azimuthal angle, formed by the spins 1 and 2.
It is interesting that the magnitude of the canting depends on $\Phi$,
which contributes to the SI anisotropy, but not to the DM energy.
Indeed, for the antiferromagnetically coupled spins, being parallel to the $[100]$ and $[110]$ axes
in the $xy$ plane,
$\cos 4 \Phi$ is equal to $1$ and $-$$1$, respectively. Therefore, in the first case (Fig.~\ref{fig.structure}c),
the effects of the SI anisotropy and DM interactions will partly compensate each other
(note that $\sin 4\phi^0_1 < 0$), while in the second case (Fig.~\ref{fig.structure}d),
these two terms will collaborate, that leads to larger spin canting. This analysis is totally consistent
with results of unrestricted HF calculations for the low-energy model (\ref{eqn.ManyBodyH}).

  The exchange coupling $J$, in the combination with the SI anisotropy $K_2'$, lifts the degeneracy
between states $\boldsymbol{e} || [100]$ and $\boldsymbol{e} || [110]$.
However, the corresponding energy difference (per one formula unit),\cite{SM}
$$
\Delta E = 2 K_2' \cos4 \phi^0_1,
$$
is very small (about $1$ $\mu$eV), mainly because $\cos4 \phi^0_1$ is small.
This is the main reason why the direction of spins in the $xy$ plane cannot be not easily
determined experimentally.\cite{Hutanu14,Soda}

  As was discussed in Sec.~\ref{sec:EPol}, the finite value of the polarization is due to the
magnetic symmetry lowering, which is reflected in the behavior of the density matrices $\hat{\cal D}_{1,2}$.
The latter can be identically presented as
$\hat{\cal D}_{1,2} = \hat{\cal D}_{1,2}^0 + \delta \hat{\cal D}_{1,2}$, where
$\hat{\cal D}_{1,2}^0$ is the average density, obeying the $P\overline{4}2_1m$ symmetry,
and $\delta \hat{\cal D}_{1,2}$ is a perturbation,
which depends on the direction of spin $\boldsymbol{e}$.
Straightforward unrestricted HF calculations yield (in $10^{-3}$)
\begin{equation}
{\rm Re} \delta \hat{\cal D}_{1,2} =
\left(
\begin{array}{ccccc}
    0  & 0  & \mp3.92  &  0  &  0  \\
    0  & \phantom{\pm}3.43 &  0  & \mp1.90 &   0 \\
   \mp3.92 &  0  &  0 &  0  & -1.30 \\
    0  & \mp1.90 &  0 &  -3.43  &  0 \\
    0  &  0  & -1.30  &  0  &  0  \\
\end{array}
\right)
\label{eqn:d12z100}
\end{equation}
and
\begin{equation}
{\rm Re} \delta \hat{\cal D}_{1,2} =
\left(
\begin{array}{ccccc}
    0  &  0  & -0.16  &  0  &  0 \\
    0  &  \pm1.85  &  0  & -2.88  &  0 \\
   -0.16  &  0  &  0 &  0  & \mp3.90 \\
    0  & -2.88 &  0  & \mp1.85  &  0 \\
    0  &  0  & \mp3.90  &  0  &  0 \\
\end{array}
\right)
\label{eqn:d12z110}
\end{equation}
for the solutions with the spin magnetization being parallel to the axes
$[100]$ and $[110]$, respectively (see Figs.~\ref{fig.structure}b and c),
where the upper (lower) signs correspond to the Co-sites 1 (2) in Fig.~\ref{fig.structure}. By combining these matrices
with $\hat{p}_{1,2z}$, given by Eq.~(\ref{eqn:p12z}), one obtains the following contributions of the sites 1 and 2
to the electric polarization: $P_{1,2z} = \pm$$31.2$ $\mu$C/m$^2$ and $P_{1,2z} = -$$28.9$ $\mu$C/m$^2$
for $\boldsymbol{e} || [100]$ and $\boldsymbol{e} || [110]$, respectively.
Thus, the magnetic structure with $\boldsymbol{e} || [100]$ preserves
the symmetry operation $\{\hat{C}_2^x | \boldsymbol{a}_1/2$$+$$\boldsymbol{a}_2/2 \}$, connecting the
sites 1 and 2, which results in the antiferroelectric state. On the other hand, the magnetic structure
with $\boldsymbol{e} || [110]$ breaks this symmetry, giving rise to the FE order.
The total polarization $P_{z} = P_{1z}$$+$$P_{2z} = 57.8$ $\mu$C/m$^2$ is in fair agreement
with the experimental data.
According to Eqs.~(\ref{eqn:p12z}), (\ref{eqn:d12z100}), and (\ref{eqn:d12z110}),
such a behavior is related to the phases of the matrix elements of
${\rm Re} \delta \hat{\cal D}_{1,2}$,
which interplay with the phases of $\hat{p}_{1,2z}$: The phases are organized in such a way that
for $\boldsymbol{e} || [100]$ and $\boldsymbol{e} || [110]$ their interplay yields
$P_{2z} = P_{1z}$ and $P_{2z} = -$$P_{1z}$, respectively. Note also that the obtained matrices
${\rm Re} \delta \hat{\cal D}_{1,2}$ do not couple with $\hat{p}_{1,2x}$ and $\hat{p}_{1,2y}$,
so that the $x$ and $y$ components of the polarization are identically equal to zero.

\begin{figure}[h!]
\begin{center}
\includegraphics[width=6cm]{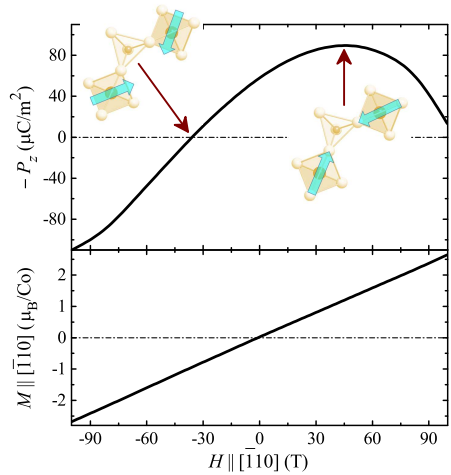}
\end{center}
\caption{\label{fig.H110}
(Color online)
Electric polarization and net spin magnetization in the external magnetic field
parallel to the $[\bar{1}10]$ axis, as obtained from the solution of the low-energy
electron model in the Hartree-Fock approximation.
}
\end{figure}

  Next, we discuss how the electric polarization can be controlled by the external magnetic field $H$.
First, we consider the situation, where the AFM spins are parallel to the $[110]$ axis
(Fig.~\ref{fig.structure}d) and apply $H$ along the perpendicular direction $[\bar{1}10]$.
The results are summarized in Fig.~\ref{fig.H110}, where for an easier comparison with experimental data
we plot $-$$P_z$.\cite{note1}
In this case, $H$ controls the magnitude of the spin canting and the direction of spins
relative to the upper (lower) O-O bonds. When both spins are parallel to the upper O-O bonds,
$-$$P_z$ takes the maximal value. The corresponding magnetic field can be easily found from the
analysis of the spin Hamiltonian, which yields
$$
\mu_{\rm B} H_m = - \frac{8J}{gS} \sin \left( \phi^0_1 - \frac{\pi}{4} \right).
$$
Using above values of $J$ and $\phi^0_1$, $g \approx 2$ and $S = 3/2$, $H_m$ can be evaluated as 47 T,
which is in the very good agreement with the results of HF calculations for the low-energy electron model
displayed in Fig.~\ref{fig.H110}. Nevertheless, $H_m$ is overestimated
by factor five in comparison with the
experimental data,\cite{Murakawa,Nakajima}
following similar overestimation of $J$, as was explained above.

  When the spins are aligned in between the upper and lower O-O bonds, $P_z$ is equal to zero.
The maximal value of the polarization at $H = H_m$ is $P_m \approx 90$ $\mu$C/m$^2$,
which is in fair agreement with the experimental data (about $120$ $\mu$C/m$^2$, Ref.~\onlinecite{Murakawa}).

\begin{figure}[h!]
\begin{center}
\includegraphics[width=8cm]{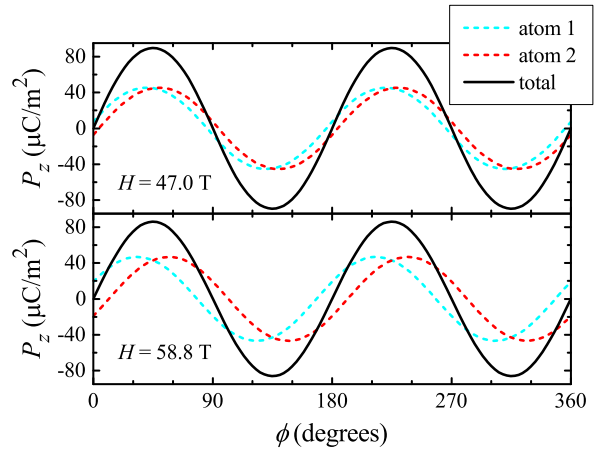}
\end{center}
\caption{\label{fig.Hxy}
(Color online)
Behavior of electric polarization (total and partial contributions of the Co sites 1 and 2)
under the rotation of the external magnetic field
in the plane $xy$, as obtained from the solution of the low-energy
electron model in the Hartree-Fock approximation. Here, the angle $\phi$ specifies the direction
of the magnetic field, while the antiferromagnetic component of the magnetization is perpendicular to the filed.
}
\end{figure}

  Since the in-plane anisotropy is small, the spins in the $xy$ plane can be easily rotated by the
magnetic field, which couples to the net magnetization. The results of such calculations
are displayed in Fig.~\ref{fig.Hxy}.
In this case, the value of the magnetic field also plays an important role,
as it controls the angle between spins at two Co sublattices: if $H \approx H_m$,
the individual contributions $P_{1z}$ and $P_{2z}$ change ``in phase'', and
for $\phi = \pi/4$ (modulo $\pi/2$)
$|P_{1z}$$+$$P_{2z}|$ reaches its maximal possible value $P_m$. However,
if $H \ne H_m$, there is some ``dephasing'' and $|P_{1z}$$+$$P_{2z}|$ is smaller than $P_m$.

\begin{figure}[h!]
\begin{center}
\includegraphics[width=10cm]{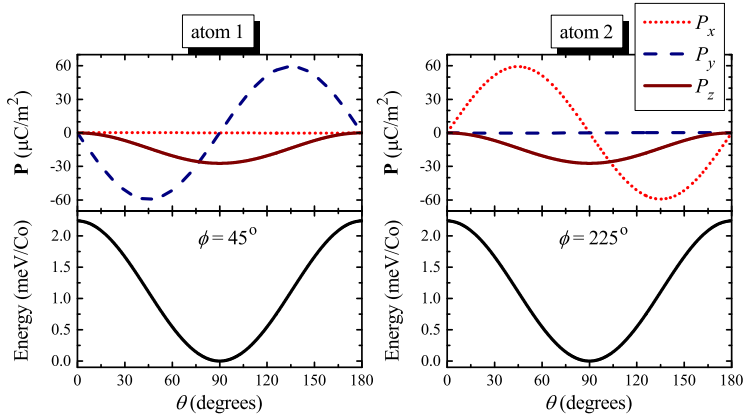}
\end{center}
\caption{\label{fig.atomicT}
(Color online)
Electronic polarization and total energy
depending on the direction
$\boldsymbol{e} = (\cos \phi \sin \theta, \sin \phi \sin \theta, \cos \theta)$ of the spin magnetization,
rotated out of the $xy$ plane for $\phi = \pi/4$ and $\phi = 5\pi/4$, as obtained in the
self-consistent linear response theory for the
spin-orbit coupling.\cite{SCLR}
The azimuthal angles $\phi = \pi/4$ and $\phi = 5\pi/4$ at the sites $1$ and $2$, respectively,
were chosen to simulate the antiferromagnetic spin alignment in the $xy$ plane.
}
\end{figure}

  As was already discussed in Sec.~\ref{sec:EPol}, for an arbitrary direction of spins,
the original $P\overline{4}2_1m$ symmetry is completely destroyed and the polarization ${\bf P}$
can also have an arbitrary direction.
Similar to the in-plane rotations (Fig.~\ref{fig.atomicP}), this behavior can be well understood already in the atomic limit,
by considering the change of the density matrix, induced by the SO interaction at a given Co site, which couples
with the electric dipole matrix $\hat{\boldsymbol{p}}_{\tau}$. For these purposes we again employ the SCLR method.
It is convenient to start with the AFM configuration of spins parallel to the $[110]$ axis
and rotate them out of the $xy$ plane. The results of such calculations are summarized in Fig.~\ref{fig.atomicT}.
One can clearly see that,
in addition to $P_z$, there are finite perpendicular components of the polarization, $P_x$ and $P_y$,
which obey certain symmetry rules and replicate the geometry of the rotated CoO$_4$ tetrahedra.
The total energy change in this case is mainly controlled by relatively large anisotropy parameter $K_1$.

\begin{figure}[h!]
\begin{center}
\includegraphics[width=8cm]{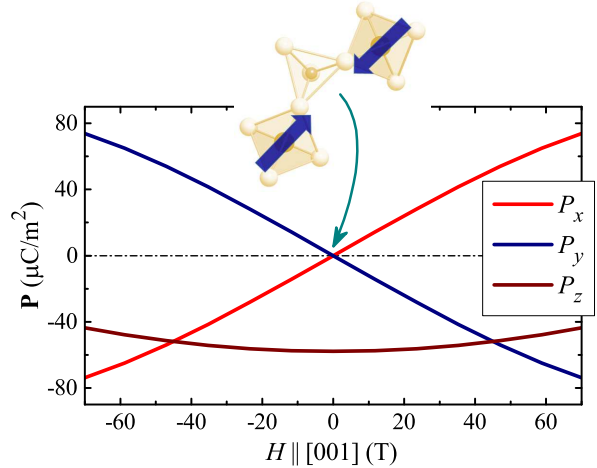}
\end{center}
\caption{\label{fig.Hz}
(Color online)
Electric polarization in the external magnetic field
parallel to the $z$ axis, as obtained from the solution of the low-energy
electron model in the Hartree-Fock approximation. The starting configuration,
where antiferromagnetic spins were parallel to the $[110]$ axis,
is shown in the inset.
}
\end{figure}

  In practice, such a situation can be realized by applying the magnetic field
along the $z$ axis, which cants the spins out of the $xy$ plane (Fig.~\ref{fig.Hz}).
The angle $\theta$ between the spins and the $xy$ plane can be estimated as
$\cos \theta = -$$\mu_{\rm B} g HS/(8J+2K_1)$. There is a clear similarity with the atomic picture,
depicted in Fig.~\ref{fig.atomicT}: the magnetic field slightly decreases $|P_z|$ and induces
two perpendicular components of the the polarization,
which satisfy the condition $P_y = -$$P_x$. According to the atomic calculations (Fig.~\ref{fig.atomicT}),
the components $P_y$ and $P_x$ are mainly induced at the Co sites 1 and 2, respectively, that is closely
related to the geometry of the rotated CoO$_4$ tetrahedra.

\section{\label{sec:Summary} Summary and conclusions}

  Using effective low-energy model, derived from the first-principles electronic structure calculations,
we have investigated the multiferroic properties of Ba$_2$CoGe$_2$O$_7$. There are two important aspects,
which make this material especially interesting in the field of multiferroic applications:
(i) Ba$_2$CoGe$_2$O$_7$ crystallizes in the noncentrosymmetric but nonpolar structure.
The key symmetry operation, which controls the multiferroic properties of this material is the rotoinversion.
From the view point of magnetic symmetry breaking,
there is a fundamental difference between inversion and rotoinversion.
The magnetic pattern, which can break the inversion symmetry, should be rather nontrivial.
Typically, such magnetic order arises from the competition of many magnetic interactions
and crucially depends on a
delicate balance between these interactions. On the contrary, the rotoinversion
symmetry can be broken very easily, simply by caning the magnetic moments out of the rotoinversion axis.
(ii) The magnetic structure of Ba$_2$CoGe$_2$O$_7$ is very soft and can be easily
deformed by the external magnetic field. This property is related to the specific geometry of
the Ba$_2$CoGe$_2$O$_7$ lattice,
where the CoO$_4$ tetrahedra are interconnected with each other via the GeO$_4$ octahedra.
Thus, the magnetic Co atoms are separated by the long Co-O-Ge-O-Co paths, resulting in the relatively
weak exchange coupling $J$. Another important ingredient is the weak in-plane anisotropy, which is
inherent to magnetic compounds with the spin $3/2$.\cite{Soda} We propose that this anisotropy has an
intraatomic nature and related to the anisotropy of the orbital magnetization.

  Nevertheless, it seems that the magnetic softness of Ba$_2$CoGe$_2$O$_7$ has also one negative aspect:
the N\'eel temperature $T_{\rm N} = 6.7$ K, below which the
multiferroic behavior has been observed, is relatively small.\cite{Hutanu11}
This imposes a serious constraint on the practical realization of the considered effects:
since $T_{\rm N}$ is controlled by the same exchange coupling $J$,
any attempts to increase $T_{\rm N}$ will make it more difficult to deform the magnetic
structure and manipulate
the properties of Ba$_2$CoGe$_2$O$_7$ by the magnetic field.

  On the theoretical side, we have shown that
the electric polarization of Ba$_2$CoGe$_2$O$_7$ can be presented as the sum of electric dipoles, which are
induced at each Co site by the local exchange field. Each such dipole is given by the expectation
value $\langle \hat{\boldsymbol{p}} \rangle = {\rm Tr} [ \hat{\boldsymbol{p}} \hat{\cal D} ]$
of the dipole matrix $\hat{\boldsymbol{p}}$ and the site-diagonal density matrix $\hat{\cal D}$.
This is rather general property of Ba$_2$CoGe$_2$O$_7$, which was derived
starting from the Berry-phase theory of electric polarization in the Wannier basis. The local character of polarization
in the case of Ba$_2$CoGe$_2$O$_7$ is directly related to the rotoinversion symmetry.

  The magnetic state dependence of the electric polarization is fully described by the
site-diagonal density matrix $\hat{\cal D}$. Any rotation of
the local magnetization from the
rotoinversion axis lowers the symmetry of the density matrix $\hat{\cal D}$ at the Co site
and induces
the local electric dipole due to the transfer of the weight of the Wannier functions to the
neighboring oxygen sites. This transfer is possible due to the $p$-$d$ hybridization.
However, the spin dependence of the hybridization itself does not play an important role.
The direction of the electric dipole depends, via $\hat{\cal D}$, on that
of the local magnetization. The total polarization of the crystal is the macroscopic average
over the microscopic electric dipoles.
This is the basic microscopic picture underlying the multiferroic behavior of Ba$_2$CoGe$_2$O$_7$.

\textit{Acknowledgements}.
This work is partly supported by the grant of Russian Science Foundation (project No. 14-12-00306).

\appendix*
\section{\label{sec:Appendix}
Polarization of oxygen band and interatomic exchange interactions}

  In this appendix, we evaluate the change of the magnetic energy, caused by the polarization of the O $2p$ band.

  After the solution of the low-energy model, consisting of the
Co $3d$ bands, we expand the basis Wannier functions of the model in the original LMTO basis,
Eq.~(\ref{eqn:overLMTO}), and construct the spin magnetization density
$m({\bf r}) = n_{\uparrow}({\bf r})$$-$$n_{\downarrow}({\bf r})$,
associated with the Co $3d$ band. In this appendix, $n_{\uparrow}$ and $n_{\downarrow}$
($v_{\uparrow}$ and $v_{\downarrow}$)
denote the electron densities (potentials) for the majority ($\uparrow$) and minority ($\downarrow$) spin states.
$m({\bf r})$ has major contributions at the Co sites as well as some hybridization-induced contribution
at the oxygen and other atomic sites. Following the philosophy of the low-energy model,\cite{review2008}
the interaction of $m({\bf r})$
with the rest of the electronic states can be described in the frameworks of
the local-spin-density approximation (LSDA). Therefore, our strategy
is to evaluate, within LSDA,
the exchange-correlation (xc) field $b({\bf r}) = v_{\downarrow}({\bf r})$$-$$v_{\uparrow}({\bf r})$,
which is induced by $m({\bf r})$ and polarizes the O $2p$ band,
and find the self-consistent change of $m({\bf r})$ and $b({\bf r})$, caused by the polarization
of the O $2p$ band.
For these purposes, it is convenient to use the SCLR theory.\cite{SCLR}
For simplicity, let us consider the discrete lattice model and assume that all weights of $m({\bf r})$
are concentrated in the lattice points:
$m({\bf r}) = \sum_{\upsilon} m_{\upsilon} \delta (\textbf{r} - \boldsymbol{\upsilon})$,
where $m_{\upsilon}$ is the local magnetic moment at the site $\upsilon$. Furthermore,
we recall that LSDA is conceptually close to the Stoner model.\cite{Gunnarsson}
Then, the magnetic part of the xc energy can be approximated as
$E_{\rm xc} = -$$\frac{1}{4} \sum_{\upsilon} I_{\upsilon} m_{\upsilon}^2$.
In practical calculations, the parameters $\{ I_{\upsilon} \}$ can be obtained using the values of
intraatomic spin splitting and local magnetic moments. In LMTO, the intraatomic exchange splitting
can be conveniently expressed via C-parameters of the centers of gravity of the
Co $3d$ states.\cite{LMTO}

  Then, by introducing the vector notations $\vec{b} \equiv [ b_{\upsilon} ]$, and
the tensors
$\hat{\cal I} = [ I_{\upsilon} \delta_{\upsilon \upsilon'}]$
and $\hat{\cal R} = [ {\cal R}_{\upsilon \upsilon'} ] $,
the self-consistent field can be found as
$$
\vec{b} = \left[ 1 + \hat{\cal I} \hat{\cal R} \right]^{-1} \vec{b}^{\,0},
$$
where
$\vec{b}^{\,0} = \hat{\cal I} \vec{m}$ is the xc field induced by the Co $3d$ band, and the response tensor
is obtained in the first order of the
perturbation theory for the wavefunctions, starting from the nonmagnetic band structure in LDA:
\begin{equation}
{\cal R}_{\upsilon \upsilon'} = \sum_{ab} \sum_n^{\rm occ} \sum_{n'}^{\rm unocc} \sum_{\bf k}^{\rm BZ}
\left\{
\frac{(C_{n {\bf k}}^{\upsilon a})^* C_{n' {\bf k}}^{\upsilon a}
(C_{n' {\bf k}}^{\upsilon' b})^* C_{n {\bf k}}^{\upsilon' b}}
{\varepsilon_{n {\bf k}} - \varepsilon_{n' {\bf k}}}
+
{\rm c.\,c.}
\right\}.
\label{eqn:Rtensor}
\end{equation}
In these notations, $\{ C_{n {\bf k}}^{\upsilon a} \}$ are the coefficients of the expansion of the LDA wave functions
over the LMTO basis, $\{ \varepsilon_{n {\bf k}} \}$ are the LDA eigenvalues, and ${\bf k}$ runs over the
first Brillouin zone (BZ). Moreover, similar to constrained random-phase approximation
for the screened Coulomb interactions (Ref.~\onlinecite{Ferdi04}),
we have to exclude from the summation in Eq.~(\ref{eqn:Rtensor}) the contributions,
where both indexes $n$ and $n'$ belong to the Co $3d$ band.
In the present perturbation theory, such
terms describe the
change of the magnetization,
which is caused by the LSDA potential in the Co $3d$ band.
However, these effects are already taken into account in the low-energy model,
where the LSDA part is replaced by a more rigorous unrestricted HF approximation
with the screened Coulomb interactions. Therefore,
such terms should be excluded
at the level of SCLR calculations for the LSDA part. In practical calculations, $n$ runs over the occupied O $2p$ bands and
$n'$ runs over the unoccupied Co $3d$ bands.

  Once $\vec{b}$ is known, the change of $\vec{m}$,
caused by the polarization of the oxygen band, can be found as
$$
\delta \vec{m} = - \hat{\cal R} \vec{b}
$$
and corresponding change of the xc-field is $\delta \vec{b} = \hat{\cal I} \delta \vec{m}$.
Since O $2p$ band is fully occupied, the net change of the magnetic moment is identically equal to zero:
$\sum_{\upsilon} \delta m_{\upsilon} = 0$, irrespectively on the type of the magnetic order.
Nevertheless, the individual moments $\delta m_{\upsilon}$ can be finite and contribute to the energy.
The corresponding correction to the total energy consists of two parts: $\delta E = \delta E_{\rm Co-O} + \delta E_{\rm O}$,
where $\delta E_{\rm Co-O} = -\frac{1}{2} \delta \vec{m}^T \hat{\cal I} \vec{m}$
is the interaction of $\delta m_{\upsilon}$ with the ``external'' xc field, created by the Co $3d$ band,
and $\delta E_{\rm O}$ is the energy change caused by $\delta \vec{m}$ in the O $2p$ band.
It also consists of two parts:
$\delta E_{\rm O} = \delta E_{\rm sp} + \delta E_{\rm dc}$, where $\delta E_{\rm sp}$ is the single-particle energy,
which can be found in the second order of $\delta \vec{b}$ as
$\delta E_{\rm sp} = \frac{1}{4} \delta \vec{b}^T \hat{\cal R} \vec{b}$,\cite{SCLR}
$\delta E_{\rm dc} = \frac{1}{4} \delta \vec{m}^T \hat{\cal I} \delta \vec{m}$ is the double-counting energy, and
$\delta \vec{m}^T$ is the row vector, corresponding to the column vector $\delta \vec{m}$.
Meanwhile, it is assumed that the magnetic energy of the Co $3d$ band itself is described by the low-energy model
in the HF approximation.

  $\delta E$ may have different values in the case of
the FM and AFM alignment of spins in the $xy$ plane. This difference additionally contribute to
interatomic exchange interactions in the plane.

  In the $P\overline{4}2_1m$ structure of Ba$_2$CoGe$_2$O$_7$, there are three types of oxygen atoms: O1, O2, and O3, which
are located in the Wyckoff positions $2c$, $4e$, and $8f$, respectively.\cite{Hutanu11}
The obtained parameters $\{ I_{\upsilon} \}$ are
$0.98$, $1.92$, $1.10$, and $2.01$ eV for Co,
O1, O2, and O3, respectively. The magnetic moments are listed in Table~\ref{tab:A1} and the energies are in
Table~\ref{tab:A2}.
\begin{table}[h!]
\caption{Local magnetic moments $m_{\upsilon}$, derived from the low-energy model for the Co $3d$ band, and
the moments $\delta m_{\upsilon}$, caused by the polarization of the O $2p$ band
in the ferromagnetic (F) and C-type antiferromagnetic (C) state of Ba$_2$CoGe$_2$O$_7$.
All magnetic moments are in $\mu_{\rm B}$ per site and the number of such sites
in the unit cell is given in the parentheses.}
\label{tab:A1}
\begin{ruledtabular}
\begin{tabular}{lcccc}
 & \multicolumn{2}{c}{F} & \multicolumn{2}{c}{C} \\
 \cline{2-3} \cline{4-5}
                & $m_{\upsilon}$  & $\delta m_{\upsilon}$
                & $m_{\upsilon}$  & $\delta m_{\upsilon}$ \\
\hline
Co ($\times 1$) & $2.251$ & $\phantom{-}0.304$  & $2.245$ & $\phantom{-}0.234$  \\
O1 ($\times 1$) & $0.005$ & $-0.004$ & $0$     & $0$      \\
O2 ($\times 2$) & $0.004$ & $-0.004$ & $0$     & $0$      \\
O3 ($\times 4$) & $0.184$ & $-0.073$ & $0.180$ & $-0.057$ \\
\end{tabular}
\end{ruledtabular}
\end{table}
\begin{table}[h!]
\caption{Magnetic contributions
to the energy of interaction between Co $3d$ and O $2p$ bands ($\delta E_{\rm Co-O}$), the single particle energy
in the O $2p$ band ($\delta E_{\rm sp}$), the total energy of the $2p$ band ($\delta E_{\rm O}$),
and the total energy ($\delta E = \delta E_{\rm Co-O} + \delta E_{\rm O}$)
as obtained for the ferromagnetic (F) and C-type antiferromagnetic (C) states.
All energies are in meV per one formula unit.
}
\label{tab:A2}
\begin{ruledtabular}
\begin{tabular}{lrr}
    & F~~~      &   C~~~ \\
\hline
  $\delta E_{\rm Co-O}$                  &  $-282.98$ & $-217.67$ \\
  $\delta E_{\rm sp}$                    &  $-6.46$ &  $-3.24$ \\
  $\delta E_{\rm O}$                     &  $27.07$ &  $16.91$ \\
  $\delta E$                             &  $-255.91$ &  $-200.76$ \\
\end{tabular}
\end{ruledtabular}
\end{table}
As expected, the moments $m_{\upsilon}$ are distributed mainly between central Co site and its neighboring sites O3.
In the FM state,
the total moment is $m_{\rm Co} + m_{\rm O1} + 2m_{\rm O2} + 4m_{\rm O3} = 3$ $\mu_{\rm B}$,
which is totally consistent with the value obtained in the Wannier basis.
The moments $m_{\upsilon}$ and $\delta m_{\upsilon}$ are parallel at the Co sites
and antiparallel at the oxygen sites. This tendency is consistent with results of the first-principles calculations
and can be explained by the hybridization between Co $3d$ and O $2p$ states.\cite{JPSJ} Therefore,
the negative sign of $\delta E_{\rm Co-O}$ is due to the contributions of the Co sites,
which are partly compensated by the positive contributions of the oxygen sites. The absolute value of $\delta E_{\rm Co-O}$
is larger for the FM state, mainly because $m_{\rm Co}$ and $\delta m_{\rm Co}$ are larger than those in the C-type AFM state.
Thus, the Co-O interaction additionally stabilizes the FM order.
Then, since the O $2p$ band is fully occupied, the contribution
$\delta E_{\rm sp}$ is relatively small. The total contribution of the O $2p$ band to the magnetic energy is positive.
This is because the
fully occupied O $2p$ band itself does not favor the magnetism and any magnetic polarization of this band increases the energy.
This also explains why
$\delta E_{\rm O}$ is slightly smaller in the C-type AFM state: the magnetic moments $\delta m_{\upsilon}$ are smaller
and, therefore, the magnetic perturbation of the O $2p$ band is also smaller. Anyway, this effect is considerably weaker in comparison
with the change of $\delta E_{\rm Co-O}$.

  In total, the magnetic polarization of the oxygen band favors the FM alignment of spins in the $xy$ plane.
By mapping the energy change $\delta E$ onto the spin model, which includes only
nearest-neighbor interactions in the $xy$ plane, the change of the exchange coupling, caused by the
polarization of the oxygen band, can be estimated as
$\delta J = \frac{1}{4}(\delta E[{\rm C}] - \delta E[{\rm F}]) \approx 14$ meV.
Thus, $\delta J$ will indeed compensate the AFM exchange coupling, obtained in the
low-energy model for isolated Co $3d$ bands.
However, despite the correct tendency, the obtained change $\delta J$ is too large
(and would lead to the FM alignment in the $xy$ plane).
There may be several reasons for it:
(i) The SCLR theory may be to crude for treating the magnetic polarization of the oxygen band
(in fact, the perturbation, which is introduced by $\vec{b}^{\,0}$ in the O $2p$ band is not small); (ii) Some quantitative estimates
may change by considering the correct crystal structure below $T_{\rm N}$ (which is not available yet).
Particularly, the original $P\overline{4}2_1m$ symmetry can be lowered by the exchange striction effects;
(iii) Tp some extent,
the correlation interactions in the Co $3d$ band, beyond the HF approximation, will additionally stabilize the
C-type AFM order. The corresponding contribution to the total energy difference between FM and C-type AFM states can be estimated
using the second-order perturbation theory (Ref.~\onlinecite{review2008}) as 2 meV per one formula unit.
This change alone does not significantly change the FM contribution $\delta J$, caused by the
polarization of the O $2p$ band. However, it can collaborate with other effects, such as the exchange striction.


\begin{thebibliography}{99}

\bibitem{KimuraTbMnO3}
T. Kimura, T. Goto, H. Shintani, K. Ishizaka, T. Arima, and Y. Tokura,
Nature \textbf{426}, 55 (2003).

\bibitem{MF_review}
T. Kimura, Annu. Rev. Mater. Res. \textbf{37}, 387 (2007);
S.-W. Cheong and M. Mostovoy,
Nature Materials \textbf{6}, 13 (2007);
D. Khomskii, Physics \textbf{2}, 20 (2009);
Y. Tokura and S. Seki,
Adv. Mater. \textbf{22}, 1554 (2010).

\bibitem{Hutanu11}
V. Hutanu, A. Sazonov, H. Murakawa, Y. Tokura, B. N\'afr\'adi, and D. Chernyshov,
Phys. Rev. B \textbf{84}, 212101 (2011).

\bibitem{Yi}
H.~T. Yi, Y.~J. Choi, S. Lee, and S.-W. Cheong,
Appl. Phys. Lett. \textbf{92}, 212904 (2008).

\bibitem{Murakawa}
H. Murakawa, Y. Onose, S. Miyahara, N. Furukawa, and Y. Tokura,
Phys. Rev. Lett. \textbf{105}, 137202 (2010).

\bibitem{Nakajima}
T. Nakajima, Y. Tokunaga, V. Kocsis, Y. Taguchi, Y. Tokura, and T. Arima,
Phys. Rev. Lett. \textbf{114}, 067201 (2015).

\bibitem{Arima}
T. Arima, J. Phys. Soc. Jpn. \textbf{76} 073702 (2007).

\bibitem{Yamauchi}
K. Yamauchi, P. Barone, and S. Picozzi,
Phys. Rev. B \textbf{84}, 165137 (2011).

\bibitem{FE_theory}
R.~D. King-Smith and D. Vanderbilt,
Phys. Rev. B \textbf{47}, 1651 (1993);
D. Vanderbilt and R.~D. King-Smith, \textit{ibid.} \textbf{48}, 4442 (1993);
R. Resta, J. Phys.: Condens. Matter \textbf{22}, 123201 (2010).

\bibitem{Barone}
P. Barone, K. Yamauchi, and S. Picozzi,
Phys. Rev. Lett. \textbf{106}, 077201 (2011).

\bibitem{DEPol}
I.~V. Solovyev and S.~A. Nikolaev,
Phys. Rev. B \textbf{87}, 144424 (2013); \textbf{90}, 184425 (2014).

\bibitem{Terakura}
S. Ishibashi and K. Terakura,
J. Phys. Soc. Jpn. \textbf{83}, 073702 (2014).

\bibitem{Nikolaev}
S.~A. Nikolaev and I.~V. Solovyev,
Phys. Rev. B \textbf{89}, 064428 (2014).

\bibitem{orth_MF}
I.~V. Solovyev, Phys. Rev. B \textbf{83}, 054404 (2011); \textbf{90}, 179910(E)
(2014); I.~V. Solovyev, M.~V. Valentyuk, and V.~V. Mazurenko, Phys.
Rev. B \textbf{86}, 144406 (2012).

\bibitem{review2008}
I.~V. Solovyev,
J. Phys.: Condens. Matter \textbf{20}, 293201 (2008).

\bibitem{WannierRevModPhys}
N. Marzari, A.~A. Mostofi, J.~R. Yates, I. Souza, and D. Vanderbilt,
Rev. Mod. Phys. \textbf{84}, 1419 (2012).

\bibitem{LMTO}
O.~K. Andersen, Phys. Rev. B \textbf{12}, 3060 (1975);
O. Gunnarsson, O. Jepsen, and O.~K. Andersen,
{\it ibid.} {\bf 27}, 7144 (1983);
O.~K. Andersen, Z. Pawlowska, and O. Jepsen, {\it ibid.} \textbf{34}, 5253 (1986).

\bibitem{Ferdi04}
F. Aryasetiawan, M. Imada, A. Georges, G. Kotliar, S. Biermann,
and A.~I. Lichtenstein,
Phys. Rev. B \textbf{70}, 195104 (2004).

\bibitem{SM}
Supplemental materials
[parameters of effective model and theoretical analysis of the spin canting in Ba$_2$CoGe$_2$O$_7$].

\bibitem{SCLR}
I.~V. Solovyev, Phys. Rev. B \textbf{90}, 024417 (2014).

\bibitem{Skomski}
R. Skomski, \textit{Simple Models of Magnetism}
(Oxford University Press, Oxford, 2008).

\bibitem{Soda}
M. Soda, M. Matsumoto, M. M{\aa}nsson, S. Ohira-Kawamura, K. Nakajima, R. Shiina, and T. Masuda,
Phys. Rev. Lett. \textbf{112}, 127205 (2014).

\bibitem{note2}
In order to be consistent with our definition of the spin model, all experimental parameters should be additionally
multiplied by $S^2$, where $S=3/2$.

\bibitem{Bruno}
P. Bruno, Phys. Rev. B \textbf{\bf 39}, 865 (1989).

\bibitem{JHeisenberg}
A.~I. Liechtenstein, M.~I. Katsnelson, V.~P. Antropov, and
V.~A. Gubanov, J. Magn. Magn. Matter. \textbf{67}, 65 (1987).

\bibitem{spinRPA}
S.~V. Tyablikov, \textit{Methods of Quantum Theory of Magnetism}
(Nauka, Moscow, 1975).

\bibitem{Hutanu14}
V. Hutanu, A.~P. Sazonov, M. Meven, G. Roth, A. Gukasov, H. Murakawa, Y. Tokura, D. Szaller,
S. Bord\'acs, I. K\'ezsm\'arki, V.~K. Guduru, L.~C.~J.~M. Peters, U. Zeitler, J. Romh\'anyi, and B. N\'afr\'adi,
Phys. Rev. B \textbf{89}, 064403 (2014).

\bibitem{note1}
Absolutely the same calculations can be performed by aligning the AFM spins parallel to the $[\bar{1}10]$
axis and applying the magnetic field along the $[110]$ axis, which will lead to the positive net polarization,
being in total agreement with the experimental data.\cite{Murakawa}

\bibitem{Gunnarsson}
O. Gunnarsson, J. Phys. F: Met. Phys. \textbf{6}, 587 (1976).

\bibitem{JPSJ}
I. Solovyev,
J. Phys. Soc. Jpn. \textbf{78}, 054710 (2009).



\end{thebibliography}
\end{document}